\documentclass[12pt]{article}
\usepackage{epsfig}
\usepackage{graphics}
\usepackage{times}
\usepackage{rotating}

\title{The Split Janus-faced Sun: Magnetic Rhythm and Duality in the Solar Cycle}
\author {WeiqI Chen$^{1, 2, 3, 4}$,  K. J. LI$^{1, 4, *}$, Jingchen Xu$^{1, 3}$    \\
\footnotesize{$^{1}$Yunnan Observatories, Chinese Academy of Sciences, Kunming 650011, China}\\
\footnotesize{$^{2}$University of Chinese Academy of Sciences, Beijing 100049, China}\\
\footnotesize{$^{3}$Yunnan Key Laboratory of Solar Physics and Space Science, 650216, China} \\
\footnotesize{$^{4}$Key Laboratory of Solar Activity, National Astronomical Observatories, CAS, Beijing 100012, China}\\
\footnotesize{$*$To whom correspondence should be addressed; E-mail: lkj@ynao.ac.cn}
}

\date{}
\begin{document}
\baselineskip24pt
\maketitle

\abstract{The solar cycle - most notably characterized by its sunspot activity patterns - serves as  a cornerstone of heliospheric physics. This research uncovers a fundamental magnetic dichotomy in the Sun's full-disk field, identifying two functionally separate populations: the Strong-Field Group (SG) and Weak-Field Group (WG). The solar cycle exhibits a dual nature, much like Janus, with the SG and WG operating in opposing phases regardless of low or high latitudes. The SG-dominated cycle represents one facet of this duality and is visually prominent at the solar surface. It is well-established that this component synchronizes with the sunspot cycle at low latitudes but operates in anti-phase at high latitudes. In contrast, the WG-driven cycle acts as its hidden counterpart, functioning in opposition to the SG at both high and low latitudes -- a behavior that had not been identified until now. Influenced by these magnetic field groups, this dual nature permeates the entire solar atmosphere, revealing that the full-disk solar activity is globally modulated by the Janus cycle. \\
{\bf keywords} Sun: magnetic fields --- Sun: activity --- Sun: atmosphere
}
\section{Introduction}           %% first-level sections will be auto-capitalized
\label{sect:intro}

The solar cycle represents a fundamental periodicity in the Sun's magnetic activity, characterized by approximately 11-year variations in sunspot numbers and related phenomena. This quasi-periodic pattern governs fluctuations in solar radiation, particle emission, the configuration of the Sun's magnetic field, and flare frequency,  occupying an extremely important position in solar physics (Hathaway 2015; Owens et al. 2022).
What drives the solar magnetic cycle? This question is ranked as the 51st among the 125 major scientific questions listed in the journal Science (2012). The magnetic field serves as the fundamental driver of solar and stellar activities (Kowalski 2024), playing a pivotal role in solar and stellar physics. The Sun is the only star that can be spatially resolved, making the investigation of the characteristics of its surface magnetic field and its interior origin  essential and instructive for understanding other stars. With advancements in spatiotemporal resolution, significant progress has been achieved in observing and researching the solar magnetic field (Jin et al. 2011; Mackay $\&$ Yeates 2012; Bellot Rubio $\&$ Orozco Suarez  2019; Petrie 2023). Furthermore, several global characteristics of the Sun's large-scale magnetic field have been identified, including the butterfly diagram illustrating sunspot latitude migration, the 22-year Hale magnetic cycle, Joy's law governing sunspot tilt angles, the anti-phase relationship between polar field (poloidal field) strength and sunspot cycle (toroidal field) progression, and the poleward drift of trailing sunspot polarities. The most fundamental of these characteristics remains the roughly  11-year solar cycle, a phenomenon frequently occurring and widely studied also in heliophysics (Owens $\&$ Forsyth 2013).
These observational findings have driven the development of dynamo theory models seeking to explain the physical mechanisms behind the solar cycle and magnetic field evolution (Babcock 1961; Leighton 1969; Charbonneau 2020a, 2020b; Karak 2023; Jiang 2024; Jiang $\&$ Zhang 2025), the mainstream model is the Babcock-Leighton type flux transport mean-field (global) dynamo (Jiang et al. 2016; Petri 2025).

On the solar surface, alongside the large-scale magnetic fields of sunspots in active regions (ARFs), there exist three main types of small-scale magnetic fields (Martin 1990; Li et al. 2022; Rempel et al. 2023): intranetwork magnetic fields (INFs), network magnetic fields (NTFs), and ephemeral-region magnetic fields (ERFs). The magnetic flux of these small-scale elements is primarily concentrated in three successive ranges: $10^{16} \sim 10^{17}$ Mx, $10^{18} \sim 10^{19}$ Mx, and $ > 10^{19}$ Mx, respectively (Parnell et al. 2009; Jin et al. 2011; Zhou et al. 2013).
Ephemeral regions are essentially micro active regions (Harvey 1992; Wang et al. 2012), and ERFs at low latitudes vary in phase with the solar cycle (Harvey 1992; Li et al. 2008; Jin et al. 2011; Jin $\&$ Wang 2014, 2015a; Li et al. 2024; Chen et al. 2025). In this study, ``low latitudes" generally refer to sunspot latitudes, while ``high latitudes" refer to the solar polar regions.
At high latitudes ($60^{\circ}$ to $90^{\circ}$), the number of filaments has been observed to be in phase with the solar cycle (Li et al. 2007). Since filaments form along neutral lines of active regions, ERFs are inferred to also be in phase with the solar cycle at high latitudes.
In contrast, NTFs are generally considered to be in anti-phase with the solar cycle at low latitudes (Meunier 2003; Jin et al. 2011; Korpi-Lagg et al. 2022; Li et al. 2024).

Using Hinode polar observations from 2008 to 2012, Shiota et al. (2012) found that at high latitudes, the average magnetic flux density of small vertical flux concentrations ($10^{15}$ ro $10^{17}$ Mx) shows no correlation with the solar cycle, whereas that of large vertical flux concentrations ($\geq 10^{18}$ Mx) varies in antiphase with it. More recently, Yang et al. (2025) extended this analysis using Hinode data from 2012 to 2021 and similarly observed that the magnetic flux of the nondominant polarity-i.e., the weak fields-remained stable.
Therefore network magnetic fields (NTFs) should be in anti-phase with the solar cycle at high latitudes - consistent with their behavior at low latitudes - and that intranetwork magnetic fields (INFs) likely have no solar-cycle dependence.
Under current observational capabilities, INFs are the weakest and smallest magnetic elements. They are also highly inclined, meaning their transverse magnetic flux exceeds their longitudinal flux (Livingston $\&$ Harvey 1975; Lites et al. 1996; Jin $\&$ Wang 2015a, 2015b; Bellot Rubio $\&$ Orozco Suarez 2019).
INFs are generally inferred to be independent of the solar cycle at all latitudes (Hagenaar et al. 2003; Buehler et al. 2013; Jin $\&$ Wang 2015a; Hanaoka $\&$ Sakurai 2020; Faurobert $\&$ Ricort 2021; Korpi-Lagg et al. 2022; Trelles Arjona et al. 2023). However, at high latitudes, long-term variations in the averaged magnetic field have been found to be in phase with the solar cycle (Trelles Arjona et al. 2023). In  addition, Figure 5(e)  of Shiota et al (2012) appears to suggest that the horizontal magnetic flux at high latitudes may also vary in phase  with the solar cycle.
The relationship between these three types of small-scale magnetic fields and the solar cycle - particularly at high latitudes - requires further investigation in the future.

Given that small-scale magnetic elements exhibit diverse relationships with the solar cycle - varying in phase, in anti-phase, or showing no correlation (Jin et al. 2011; Li et al. 2024; Chen et al. 2025) - the prevailing view is that local dynamos are responsible for generating these small-scale magnetic fields (Bellot Rubio $\&$ Orozco Suarez 2019; Cliver et al. 2024).
Under the influence of magnetic fields, some solar activity phenomena have been found to vary in phase with the solar cycle, others in anti-phase, and still others show no clear relationship (Jin et al. 2011; Shokri et al. 2022). Even the same phenomenon, such as X-ray bright points, has been reported by different researchers to exhibit different solar-cycle relationships  (Davis 1983; Hara $\&$ Nakakubo-Morimoto 2003), depending on the types of magnetic fields involved.
It is therefore important to understand the solar-cycle behavior of different magnetic element types and their relationship with various solar activity phenomena - a topic that merits further detailed investigation.

Full-disk magnetic field observations have been conducted for nearly half a century, serving as a critical foundation for dynamo models. However, limited research has been devoted to the detailed analysis  of these observations (Jin et al. 2011; Jin $\&$ Wang  2012; Lites et al 2014). Through an analysis of high-precision magnetograms (limited to within $60^{\circ}$ of the solar disk center), Jin et al. (2011) found that small-scale magnetic fields at low latitudes exhibit distinct solar cycle variations compared to the large-scale magnetic fields of sunspots.
Thus, it is imperative to investigate the types of magnetic fields present across the Sun's entire surface, through the analysis of long-term full-disk magnetic field observations. Additionally, achieving a more profound understanding of the polar magnetic field holds substantial significance. In order to enrich our understanding of the solar activity cycle,
we will subsequently focus on investigating the long-term evolution of the Sun's magnetic field across the solar disk, with particular focus on polar field dynamics and their role in solar cycle behavior, through analyzing full-disk observational data of the solar surface magnetic field.

% Authors can give a citation as `\citealt{Michel+etal+1992}'.
% You may also use \cite, \citep and \citet for citation, and use Table~1
% or Figure~1 and so forth. Using \ref and \label for cross-references of
% Tables/Figures is a good way in adjusting/adding/removing text, tables or
% figures.

\section{Data}
The data used in this study consist of synoptic magnetograms from Carrington rotations 1625 to 2007 (from 1975 Feb. 19  to 2003 Sept. 25), obtained by the vacuum telescope at the National Solar Observatory (NSO/Kitt Peak). These magnetograms are shown in Figure 1. Unless otherwise specified, the magnetic field strength is measured in gauss (G).
The daily sunspot number (version 2) within the same period as the magnetograms comes from the Sunspot Index and Long-term Solar Observations, which is used to represent the phase of the solar cycles and shown here in Figure 2.
In solar physics, high and low latitudes are typically delineated at approximately $55^{\circ}$ on the solar disk (Yang et al. 2024a). Here, low latitudes are quantitatively defined as the range $0 - 50^{\circ}$, while high latitudes span $60^{\circ} - 90^{\circ}$. The intermediate latitude range ($50^{\circ} - 60^{\circ}$) is excluded from this analysis.

\begin{figure}
   \centering
  \includegraphics[width=12cm]{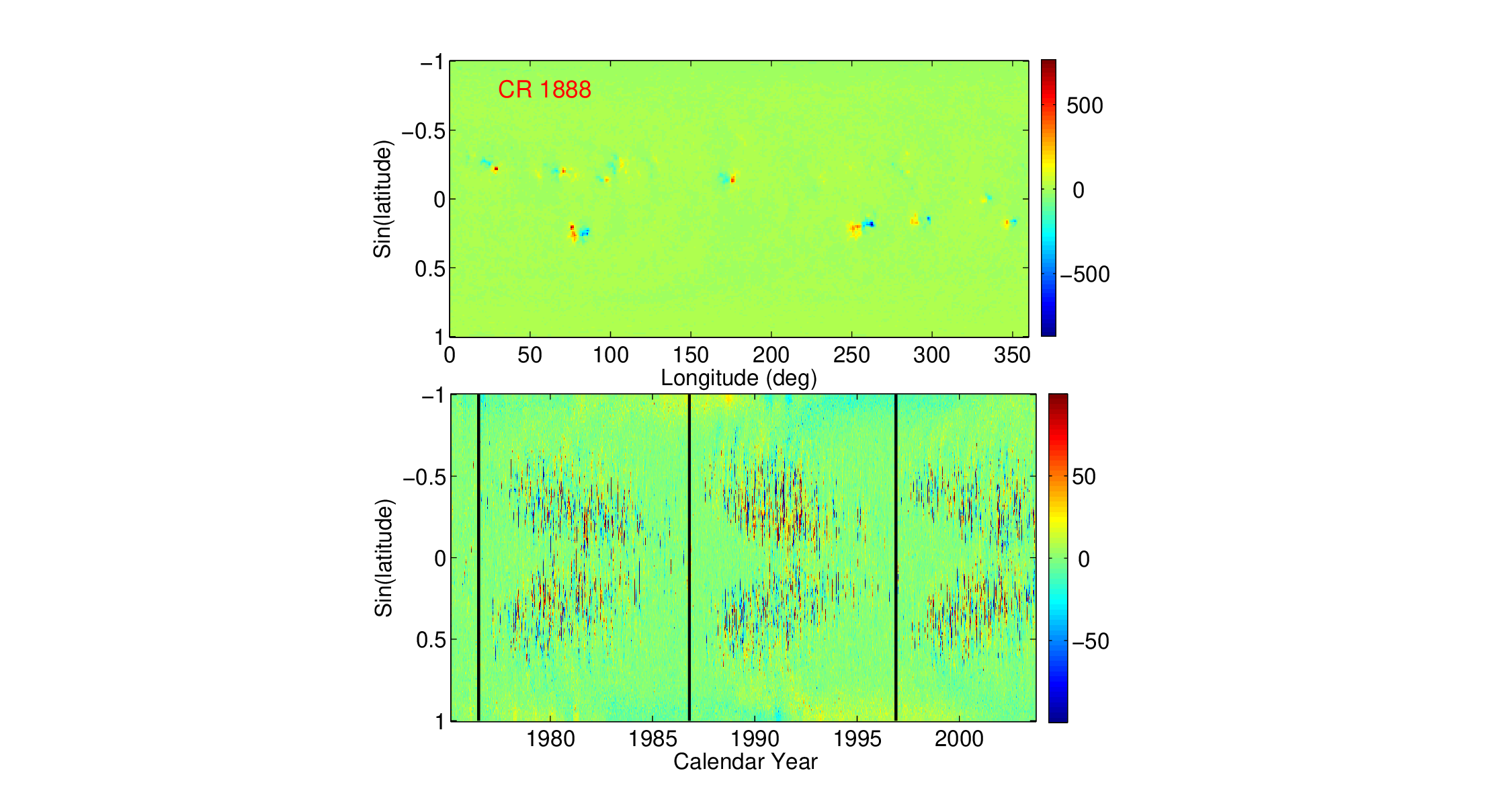}
   \caption{Top panel: the synoptic magnetogram of Carrington rotation (CR) 1888.  Bottom panel: synoptic magnetograms from CRs 1625 to  2007, with the color scale saturated at 100 G. The vertical black lines indicate the minimum epochs of the solar cycles. In the mid- and low-latitude regions, the butterfly pattern is distinctly observed.
   }
   \label{Fig2}
   \end{figure}
\begin{figure}
   \centering
  \includegraphics[width=12cm]{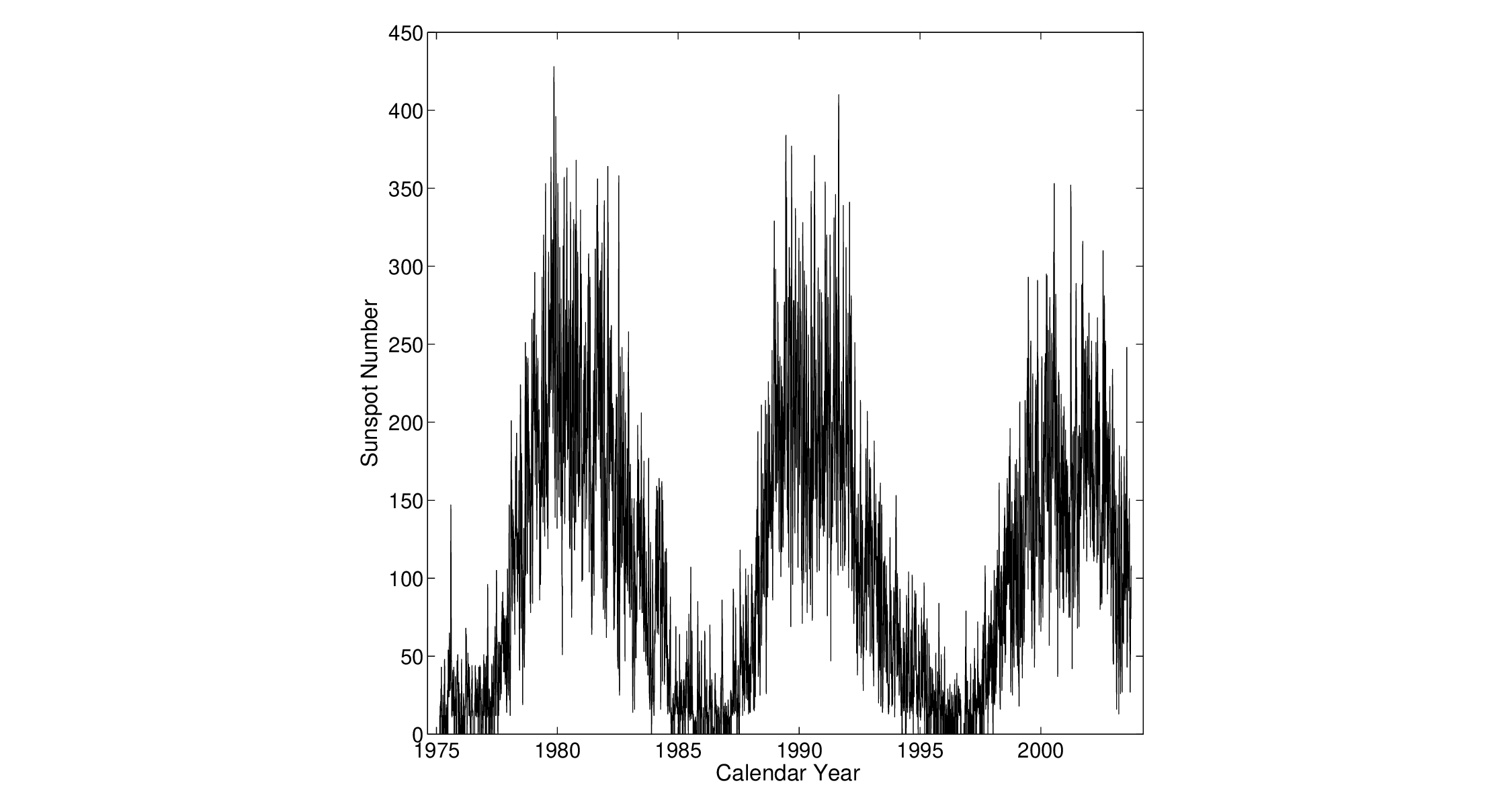}
   \caption{Daily sunspot number (version 2) from 1975 Feb. 19 to 2003 Sept. 25.
   }
   \label{Fig2}
   \end{figure}

\section{Results}
\subsection{Duality in the Solar Cycle Viewed from High and Low Latitudes}
At low latitudes, the field strength was binned into 150 evenly spaced intervals from 0 - 100 G, with values above 100 G (1.17$\%$ of all data at low latitudes) treated as 100 G. At high latitudes, 45 intervals were used from 0 - 30 G, with values above 30 G (0.42$\%$ of all data at high latitudes) treated as 30 G.
Then, a correlation analysis (Figure 3) between daily total magnetic field strength values for each interval - measured respectively at low (0 - 50$^{\circ}$) and high (60$^{\circ}$ - 90$^{\circ}$) latitudes in synoptic magnetograms - and daily sunspot numbers reveals that magnetic elements across the solar surface can be grouped into four distinct components, as summarized in Table 1 and illustrated in Figure 4. As the figure clearly shows, these four magnetic field components present different phases of the cycle at high and low latitudes.
In the above analysis, magnetic field strength values greater than 100 G are considered as 100 G, which account for $\sim 1.1\%$ of the total data at low latitudes, and  additionally, it makes no distinction between the northern and southern hemispheres.
The $95\%$ confidence level lines are shown in Figure 3.
Due to the large number of data points used in each calculation (no less than 5000), the tabulated/critical correlation coefficients are all near 0.062. As a result, the confidence level line is essentially horizontal at this value.
Except for the transition zones between components, the obtained Pearson correlation coefficients are statistically significant across nearly the entire strength range considered.
In the table, the magnetic field strength value of 5.29 represents the intersection point of the two correlation coefficient curves in Figure 3, while the value of 18.37 corresponds to the magnetic field strength at which the correlation coefficient becomes zero in the high-latitude curve shown in Figure 3. Together with a third value of 80, these three values divide the magnetic field strength into four distinct components (see Table 1 and Figure 4).

Substituting strength values of magnetic elements with their  counts  yields results that remain virtually unchanged.
At both high and low latitudes, magnetic elements in these components exhibit distinct solar cycle phase correlations, dependent on their magnetic field strengths. The critical distinction between Components III and  IV lies in their scales: Component IV consists predominantly of large-scale sunspot-associated magnetic fields, while Component III comprises small-scale magnetic elements distributed ubiquitously across the solar surface. Of course, this is an approximate division due to the existence of transition.

\begin{figure}
\centerline{\includegraphics[width=11cm, angle=0]{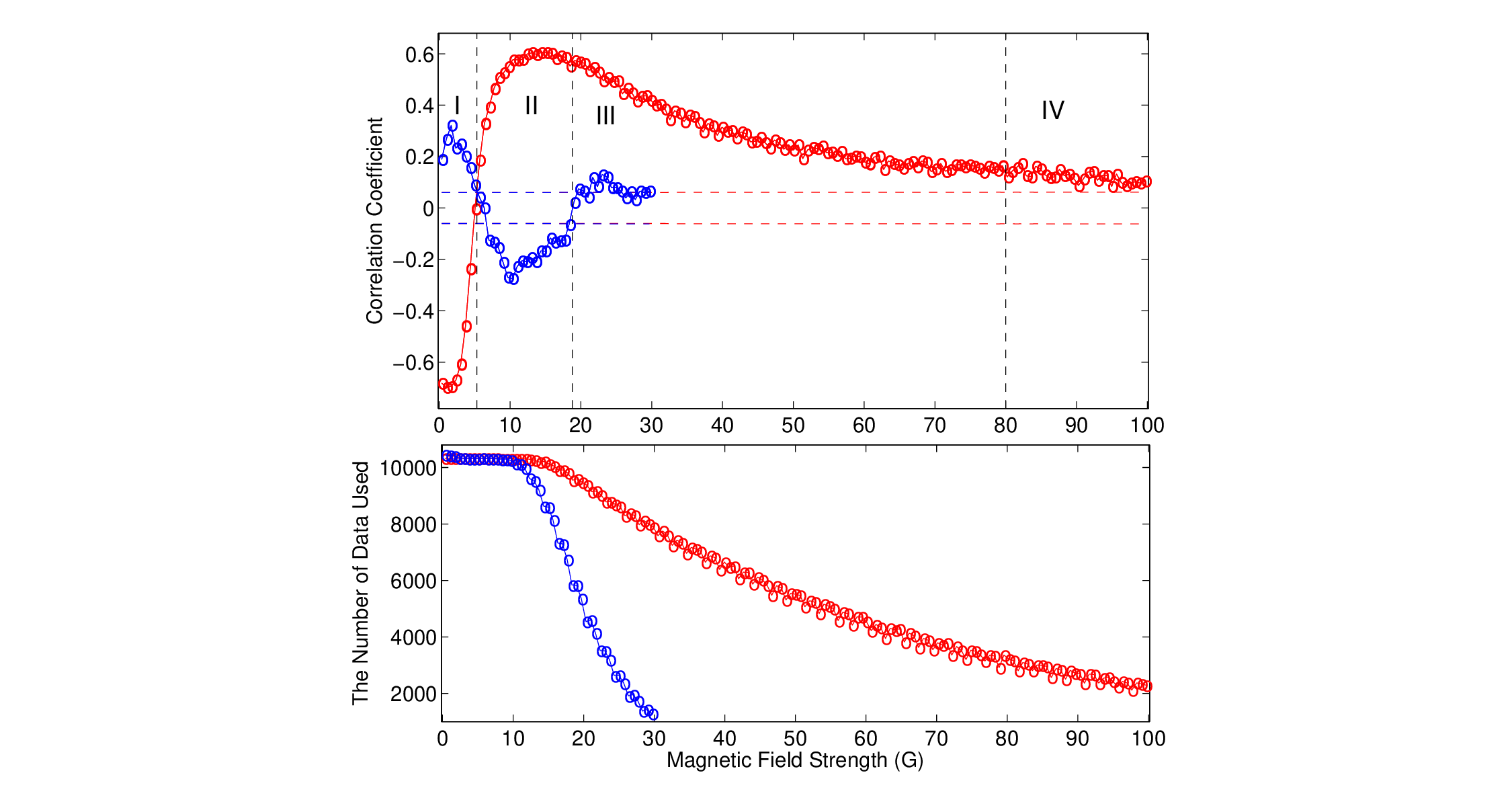}}
\small		\caption{Top Panel: Pearson correlation coefficients between daily magnetic field strength values (from magnetograms) within distinct strength ranges and daily sunspot numbers, calculated separately for low-latitude (red circles) and high-latitude (blue circles) regions. The almost horizontal red and blue dashed lines denote the $95\%$ confidence levels for their respective correlations. Vertical black dashed lines divide the magnetic elements across the solar disk into four field-strength regimes (labeled I, II, III, and IV), corresponding to weak-to-strong magnetic strength ranges.
		Bottom Panel: Number of data points (red and blue circles) used to compute the correlations in the top panel for low- and high-latitude regions, respectively. Sample sizes may vary due to differences in the frequency of magnetic elements across field-strength regimes (stronger magnetic fields are typically less frequent on the solar surface).
	}\label{}
\end{figure}

\begin{figure}
\centerline{\includegraphics[width=11cm, angle=0]{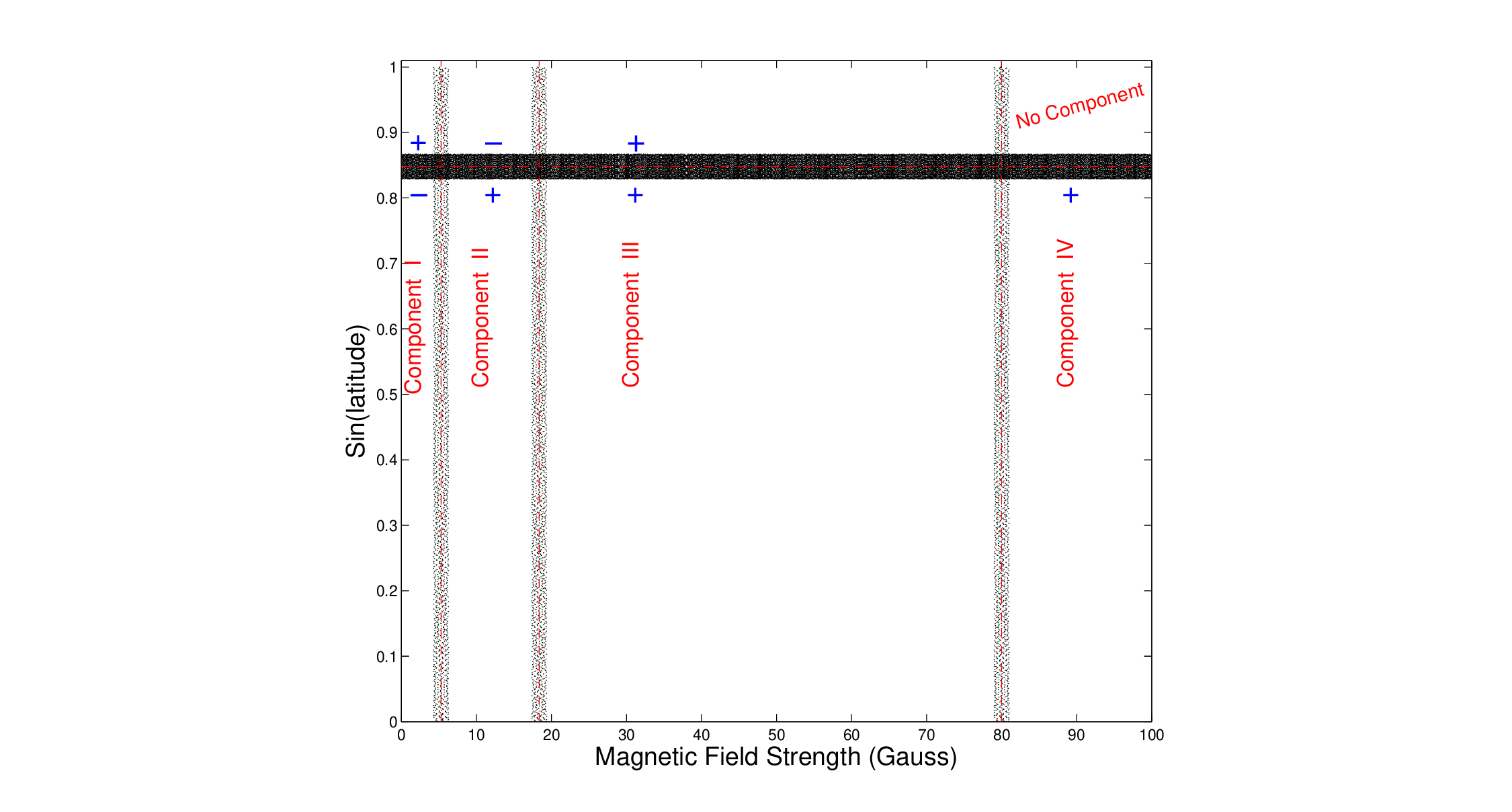}}
\small		\caption{Schematic diagram of the division (vertical red lines) of the solar magnetic field into four components (Components I to IV). The horizontal red dashed line divides the solar surface into high and low latitudes. Shadow areas indicate approximate divisions. The blue plus sign indicates being in phase with the solar cycle, while the blue minus sign, being in phase with the solar cycle.
	}\label{}
\end{figure}

For each component, we calculated its correlation coefficient with daily sunspot numbers and tabulated the results. Additionally, we analyzed the magnetic element count distributions and their cumulative distribution percentages at low and high latitudes (Figures 5 and 6). Based on these distributions, the proportional contribution of each component was quantified and listed in Table 1. Here, cumulative distribution percentage is defined as the percentage of data points at or below a specific intensity threshold relative to the total dataset.

\begin{figure}
\centerline{\includegraphics[width=11cm, angle=0]{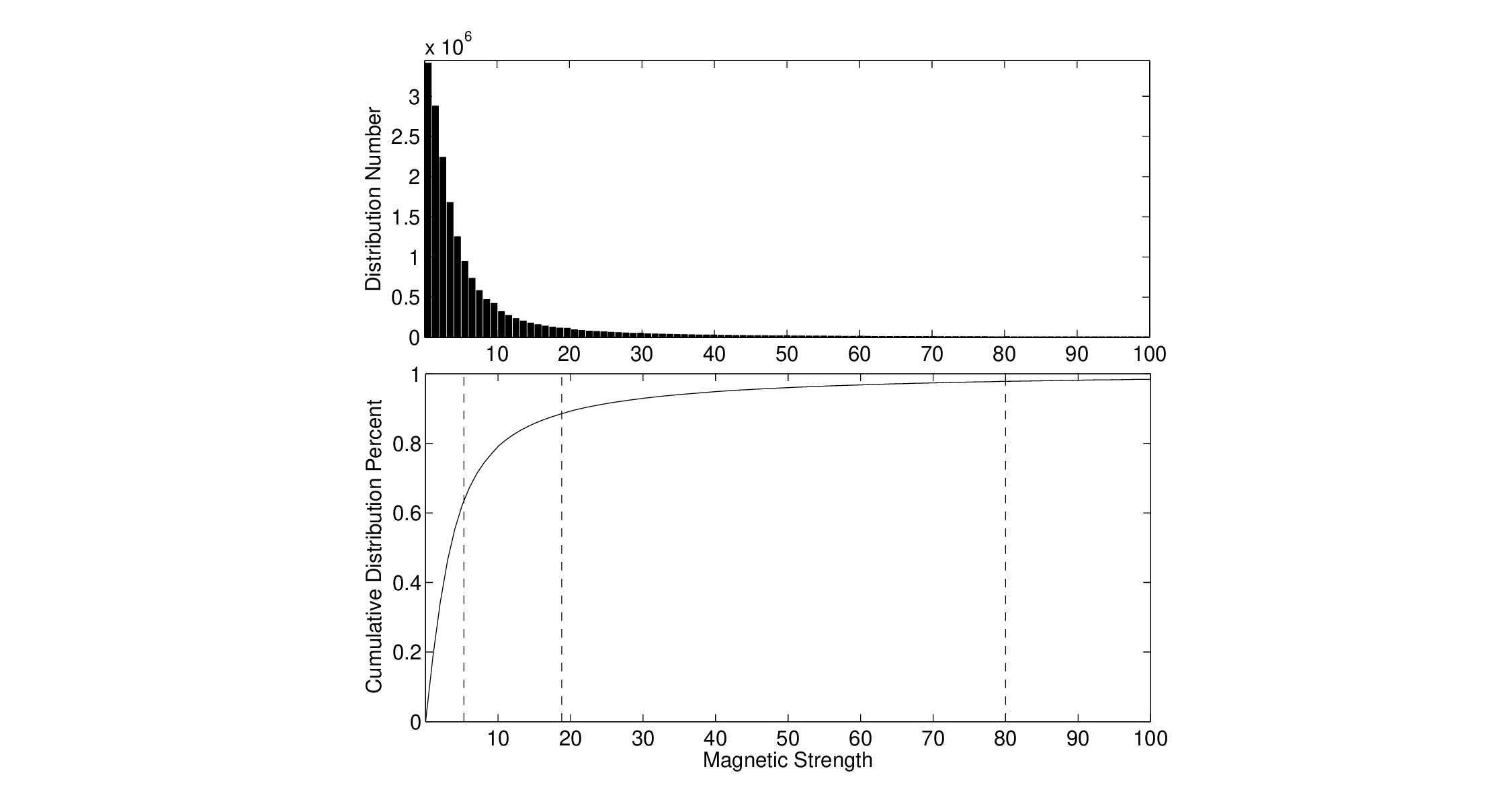}}
\small			\caption{The distribution histogram (upper panel) and the cumulative distribution percentage (lower panel) of magnetic element counts at low latitudes in synoptic magnetograms from Carrington rotations 1625 to 2007. The vertical dashed lines correspond to magnetic field strength values of 5.29 G, 18.37 G, and 80 G, which categorize the magnetic field into four distinct components.}\label{}
 \end{figure}

\begin{figure}
\centerline{\includegraphics[width=11cm, angle=0]{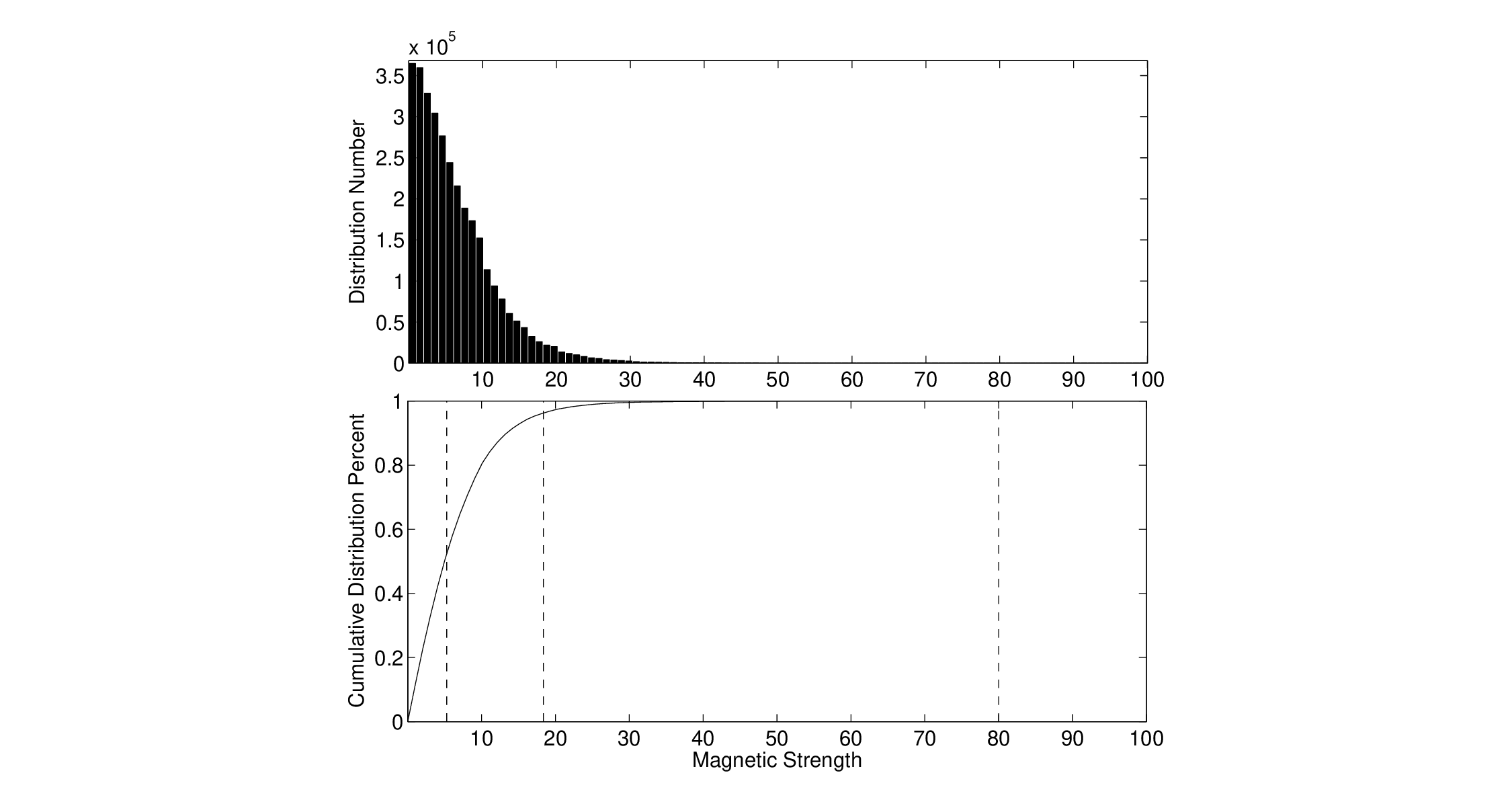}}
\small			\caption{The same as  Figure 5, but high latitudes replace low latitudes.
		}\label{}
 \end{figure}

%\lipsum[1-2]
%\vskip 2.0 cm
\newpage
\begin{sidewaystable}
%\begin{table*}
		\centering
	\caption{Characteristics of the four - component classification of magnetic elements across the full disk.}
	\begin{tabular}{lllllllll}
		\hline
		& \multicolumn{4}{l}{Low latitudes} & \multicolumn{4}{l}{High latitudes}  \\
		\cline{2-9}  \\
		Component  & I & II & III & IV & I & II & III & IV \\
		Range of magnetic field strength   &  0\,--\,5.29 & 5.29\,--\,18.37 & 18.37\,--\, 80 & $>$80 &  0\,--\,5.29 & 5.29\,--\,18.37 & 18.37\,--\, 80 & $>$80   \\
		Ratio in terms of strength  & $12.5\%$ & $22.1\%$ & $31.8\%$ & $33.6\%$ & $20.5\%$ & $65.4\%$ & $14.1\%$ & $0.08\%$   \\
		Ratio in terms of number & $63.5\%$ & $24.8\%$ & $9.5\%$ & $2.2\%$ & $52.3\%$ & $43.8\%$ & $3.8\%$ & $0.06\%$  \\
		Correlation coefficient*  & -0.709 &   0.642 &   0.711 &   0.410 &  0.389  & -0.379 & -0.041 & \\
		Correlation coefficient** &-0.626  &   0.670 &   0.700 &   0.364 &  0.358  & -0.392 & -0.040 &  \\
		Relation with the solar cycle &  anti-phase & in-phase & in-phase & in-phase  & in-phase & anti-phase & no relation  & Inconsiderable   \\
		\hline
	\end{tabular} \\
	$^{*}$When analyzing the daily total number of magnetic elements \\
	$^{**}$When analyzing the daily total magnetic field strength  \\
%\end{table*}
\end{sidewaystable}

%\vspace{6mm}
\begin{table}
%\newpage
%\begin{sidewaystable}
		\centering
	\caption{Characteristics of the weak and strong magnetic groups.}
	\begin{tabular}{lllll}
		\hline
		& \multicolumn{2}{c}{Low latitudes} & \multicolumn{2}{c}{High latitudes} \\ \cline{2-5} \\
		Group  & WG  & SG  & WG  & SG   \\
		Range of magnetic field strength   &  0\,--\,5.29 & $>$5.29 & 0\,--\,5.29 & $>$5.29  \\
		Ratio in terms of strength & $12.5\%$ & $87.5\%$ & $20.5\%$ & $79.5\%$ \\
		Ratio in terms of number   & $63.5\%$ & $36.5\%$ & $52.3\%$ & $47.7\%$  \\
		Correlation coefficient*   &  -0.709  & +0.728 & +0.389 & -0.362   \\
		Correlation coefficient**  &  -0.626  & +0.633 & +0.358 & -0.317    \\
		\hline
	\end{tabular} \\
	$^{*}$When analyzing the daily total number of magnetic elements \\
	$^{**}$When analyzing the daily total magnetic field strength  \\
\end{table}
%\end{sidewaystable}

As Table 1 shows, at high latitudes, Components III and IV collectively account for less than $4\%$ of magnetic elements - far smaller than the proportion ($43.8\%$) of Component II. This supports a streamlined, two-group classification of magnetic elements: a Weak magnetic Group (WG) and a Strong magnetic Group (SG). Specifically, magnetic elements with field strengths $\leq$ 5.29 G are classified as WG (corresponding to Component I), while those exceeding 5.29 G are classified as SG (encompassing Components II, III, and IV), and such a  classification is consistent across both high and low latitudes. We calculated the proportional distribution of WG and SG at both high and low latitudes. Additionally, we determined the correlation coefficients between sunspot numbers and two parameters for each group: (1) the daily total magnetic field strength and (2) the daily number of magnetic elements. These results are summarized in Table 2 and illustrated in Figure 7. The correlation coefficient values presented in Table 2 are calculated based on more than 10,000 data points, with all values demonstrating statistical significance even at the $99.9\%$ confidence level (p $<$ 0.001).
The SG and WG exhibit nearly identical correlation coefficients with the solar cycle, and thus the phase relationship of WG with the solar cycle exhibits statistical reliability comparable to that of SG, with both demonstrating robust statistical reliability.

\begin{figure}
\centerline{\includegraphics[width=11cm, angle=0]{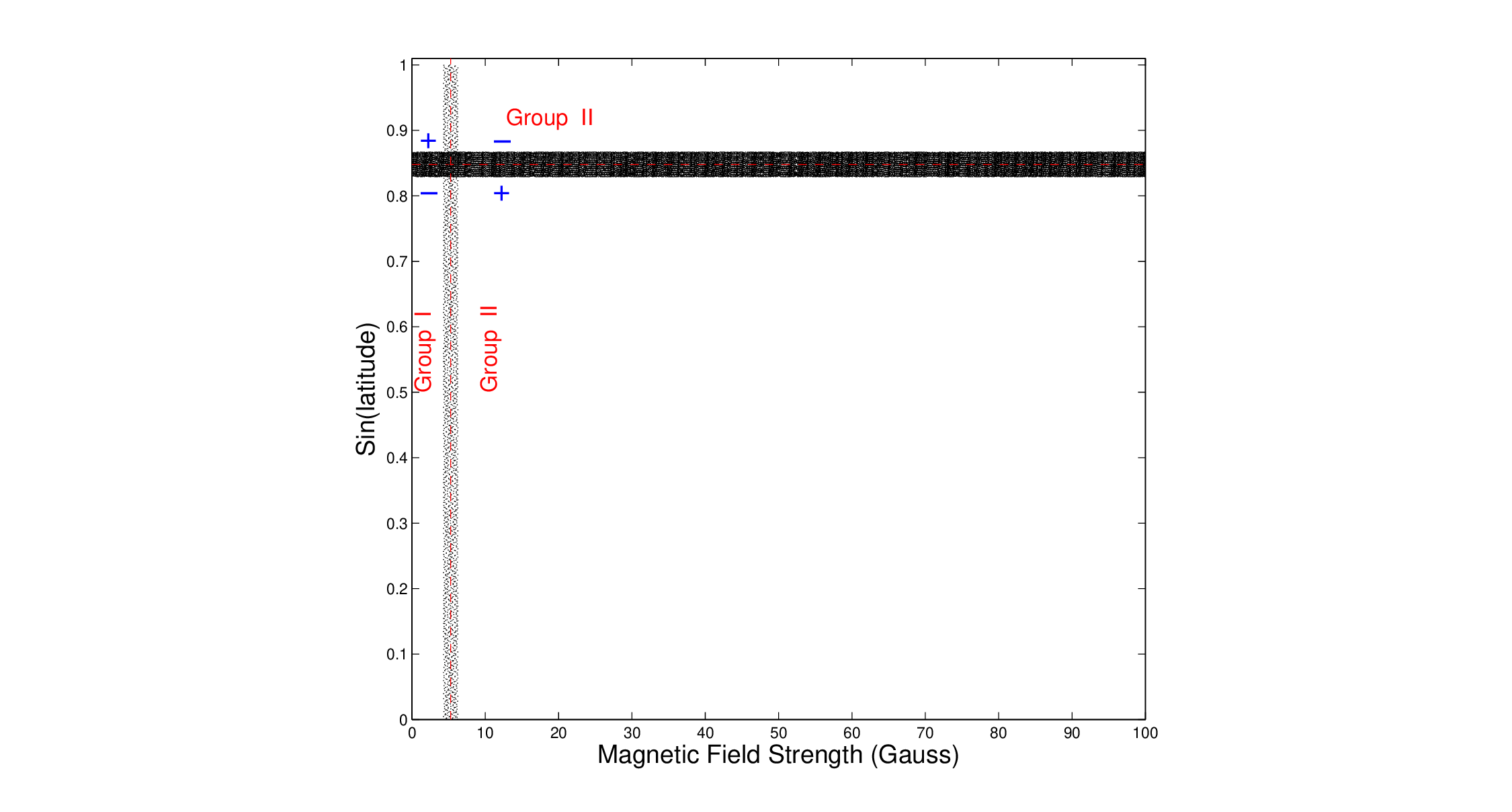}}
\small		\caption{Schematic diagram of the division (vertical red lines) of the solar magnetic field into two groups (Groups I and II). The meanings of each symbol are given in Figure 4.
	}\label{}
\end{figure}

Intriguingly, as illustrated in Figure 7, full-disk magnetic elements bifurcate into two distinct groups (SG and WG) at a threshold of approximately 5.29 G, unveiling a fundamental magnetic dichotomy. Solar magnetism exhibits a Janus-like duality, mirroring the two-faced Roman god: SG manifests as the overt component, while WG constitutes its hidden counterpart.
At low latitudes, the strong magnetic group (SG) dominates, displaying long-term evolution tightly coupled with the solar cycle. Conversely, at high latitudes, the predominant magnetic ``face"  follows an anti-solar-cycle pattern-again governed primarily by SG. This result is consistent with the findings of Yang et al. (2024b) shown in their Figure 3, where the average flux density, contributed by the predominant polarity, reaches its peak at solar minimum.
However, concealed beneath this dominant SG signature lies the weak magnetic group (WG), which behaves in precise opposition to SG. Strikingly, WG shows an anti-solar-cycle long-term pattern at low latitudes while synchronizing tightly with the solar cycle at high latitudes -presenting an inverted mirror image of SG's behavior across hemispheric domains.
In conclusion, the full-disk magnetic activity of the Sun exhibits a Janus-faced cycle.

\subsection{Duality in the Solar Cycle  Viewed from a Measurement  Latitudes}
We tallied the counts of both groups and their combined totals (i.e., all magnetic elements) across 180 measurement latitudes and calculated their correlation coefficients with daily sunspot numbers. The results are shown in Figure 8.
At each of lower measurement latitudes ($\le 56^{\circ}$), Group I is out of phase with the solar cycle, while Group II is in phase with it. Conversely, at each of higher latitudes ($\ge 62^{\circ}$), Group I is in phase with the solar cycle, while Group II is out of phase. The only exceptions occur at a few latitudes near the northern pole, likely due to relatively lower accuracy of high-latitude measurements.
When the Sun's magnetic field is treated as binary, the solar cycle's influence on the magnetic field becomes pronounced-evident even across different measurement latitudes. Thus, comprehensive observations from both high and low latitudes (and even at varying latitudes) reveal that the Sun's full-disk activity exhibits a Janus-faced cycle.
Due to this opposing cycle behavior of Groups I and II, their combined counts only weakly reflect the solar cycle - and only within the latitude range of the sunspot butterfly diagram.
By the way, a more appropriate division between high and low latitudes is $\sim 59^{\circ}$ (slightly different from the traditional $55^{\circ}$), with a transition zone of $\sim 3^{\circ}$.

\begin{figure}
\centerline{\includegraphics[width=11cm, angle=0]{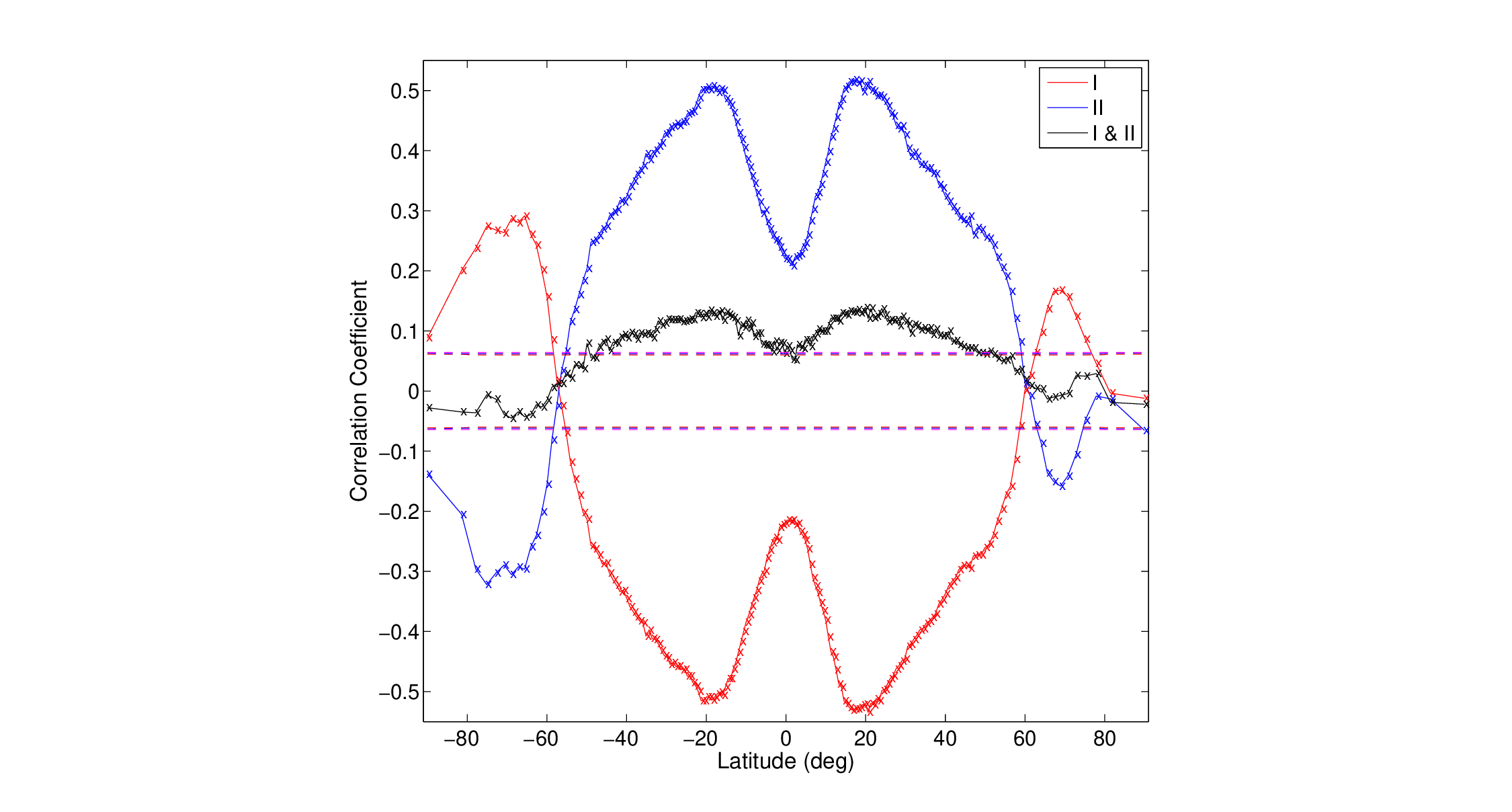}}
\small		\caption{Correlation coefficients between sunspot numbers and the daily counts of Group I (red crosses, red solid line), Group II (blue crosses, blue solid line), and their combined totals (black crosses, black solid line), across 180 measurement latitudes. The dashed lines (color-matched) represent the $95\%$ confidence intervals for the respective correlation coefficients.
	}\label{}
\end{figure}

Figure 9 displays the number of data points used to compute the correlation coefficients shown in Figure 8 and the subsequent Figure 10. Based on the sample sizes in Figure 9, the $95\%$ confidence levels were determined (dashed lines in Figsure 8 and 10).
As evident in Figure 9, the daily occurrence rate of Group I is significantly higher than Group II at low latitudes, but this trend reverses at high latitudes, where Group I becomes less frequent than Group II.

\begin{figure}
\centerline{\includegraphics[width=11cm, angle=0]{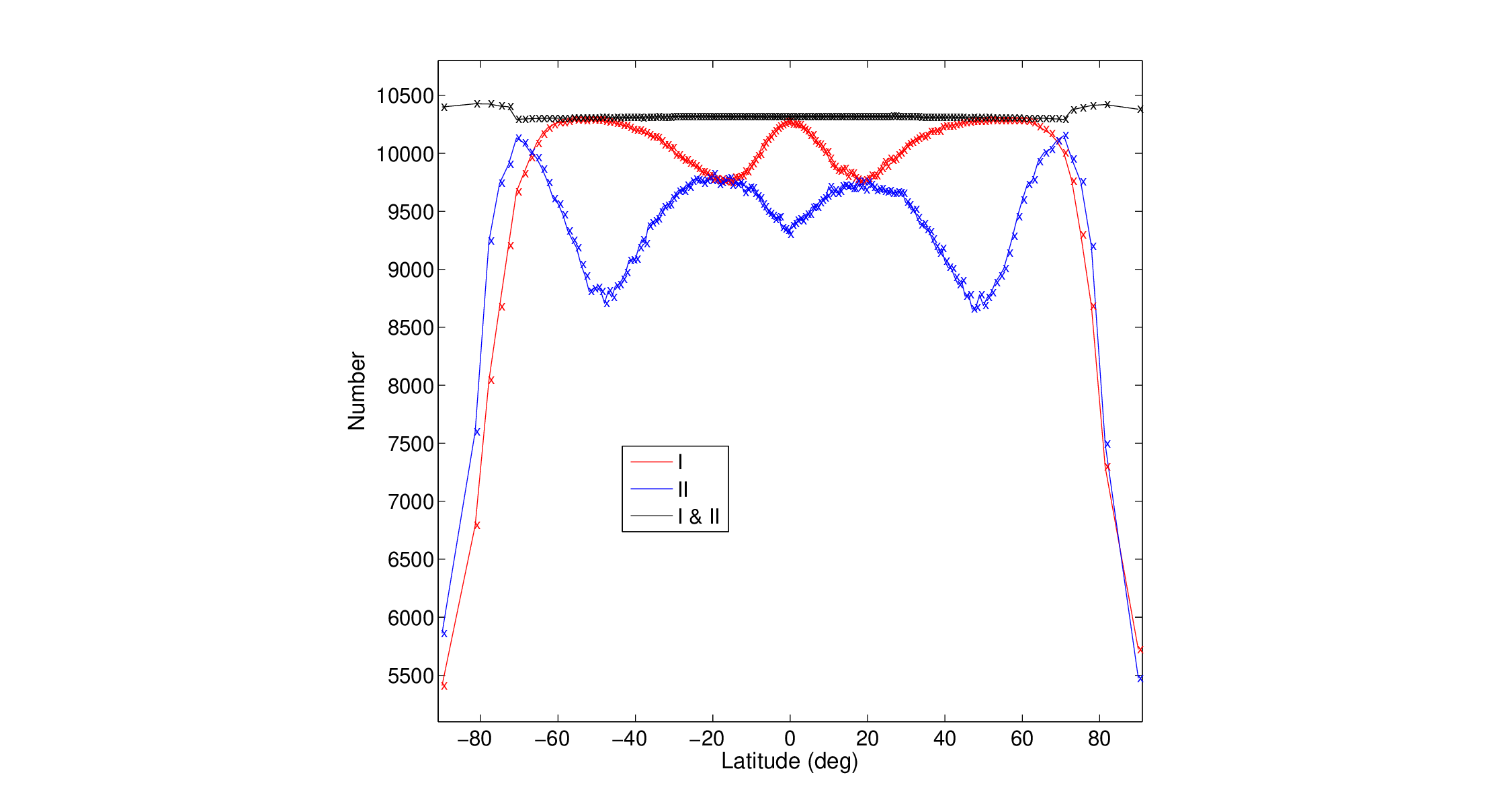}}
\small		\caption{The number of data points used to compute the correlation coefficients (color-matched) shown in Figures 8 and 10.
	}\label{}
\end{figure}

Following the same methodology, we calculated the daily magnetic field strengths for both groups (I and II) along with their combined total strength (representing all magnetic elements) across 180 measurement latitudes. The correlation coefficients between these magnetic strength measurements and sunspot numbers are presented in Figure 10. These results show essentially the same pattern as those observed in Figure 8.

Although Groups I and II exhibit opposite-phase periodic behavior, the combined results (Group II plus I) closely match those of Group II alone-except at several polar latitudes. This indicates that the daily magnetic field strength of group II is overall significantly dominant over Group I, effectively masking the latter's periodic signal in most regions. Previous observations based on magnetic intensity alone revealed only the dominant signature of the Janus cycle (Group II), traditionally identified (well-known) as the solar activity cycle. However, our analysis of magnetic element counts demonstrates that Group I occurs with significantly higher frequency than Group II at low latitudes (see Figure 9).

\begin{figure}
\centerline{\includegraphics[width=11cm, angle=0]{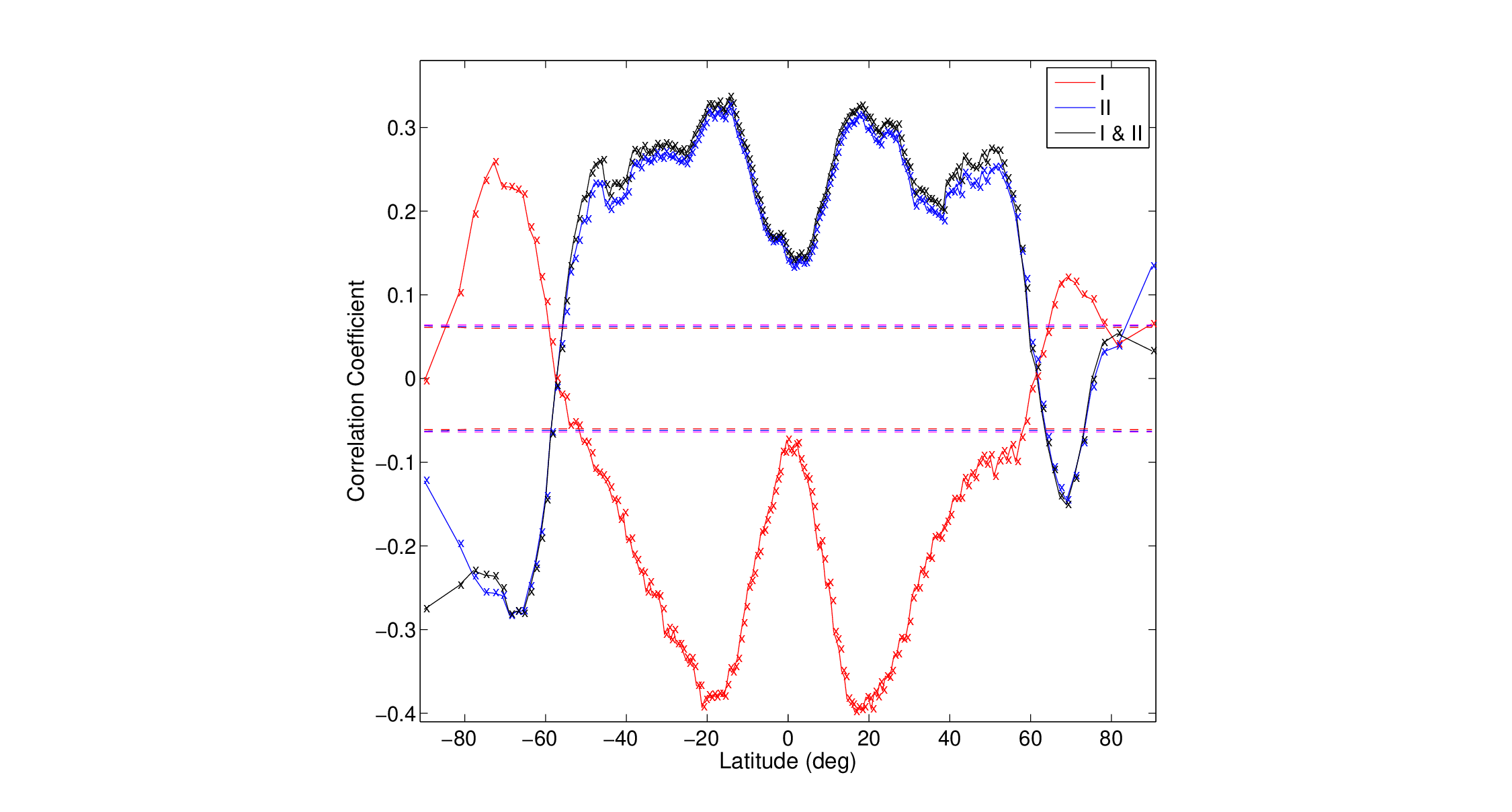}}
\small		\caption{Same as Figure 8, but tallying magnetic element strength rather than magnetic element count.
	}\label{}
\end{figure}

 \section{Conclusions and Discussion}
This study of the Sun's synoptic magnetograms  reveals a fundamental magnetic dichotomy, with 5.29 G serving as the critical boundary separating distinct solar cycle behaviors: the strong magnetic field group (SG) and the weak magnetic field group (WG).
The prominent SG exhibits the conventional pattern: in-phase synchronization with the solar cycle at low latitudes, but anti-phase behavior at high latitudes.
Remarkably, when we only consider the weak magnetic field group (WG) and ignore the strong magnetic field group (WG), a hidden aspect emerges: WG manifests out-of-phase behavior at low latitudes while maintaining in-phase synchronization with the solar  cycle at high latitudes.
This reveals the Sun's magnetic field as having two complementary faces - the familiar SG pattern and its subtle WG counterpart.
Such the systematic opposition between SG and WG - where each group exhibits an inverse solar cycle relationship relative to the other - produces what we term the ``Janus cycle" of solar full-disk activity. The phenomenon manifests as two antithetical solar cycle signatures coexisting simultaneously across all elements in synoptic magnetograms, mirroring the dual nature of the two-faced Roman god Janus.

Under the influence of solar magnetic fields, duality behavior propagates from the magnetic field in the photosphere throughout the solar atmosphere. At low latitudes, long-term aggregate brightness variations in the corona, the chromosphere, and even the photosphere align in phase with the solar cycle, while their background brightness displays an anti-phase relationship (Li $\&$ Feng 2022; Li 2024; Chen et al. 2025). Notably, total solar irradiance (TSI) may serve as a proxy for photospheric aggregate brightness.
Polar brightening exhibits an atmospheric divide: in the transition region and corona, it peaks during solar maximum, whereas photospheric and chromospheric polar brightening dominates during solar minimum. This implies that at high latitudes, the transition region and corona exhibit in-phase synchronization with the solar cycle, while the chromosphere and photosphere show anti-phase behavior (Li $\&$ Feng 2022; Li 2024; Chen et al. 2025). An intriguing phenomenon emerges in solar atmospheric dynamics at high latitudes: weak magnetic fields predominantly influence the transition region and corona, while strong magnetic fields exert their primary effects in the photosphere and chromosphere.
Critically, it remains unclear whether high-latitude atmospheric layers harbor a background brightness component inversely correlated with their aggregate brightness. Resolving this uncertainty demands high-precision polar observational data, particularly for disentangling overlapping signals across atmospheric strata.

Brightness variations correspond to atmospheric heating processes. At low latitudes, the strong magnetic field group (SG) dominates heating across the solar atmosphere, including the anomalously heated chromosphere and corona. The weak magnetic field group (WG) also contributes to global atmospheric heating, but in a subtle way.  At high latitudes, heating mechanisms diverge: the photosphere and chromosphere are primarily heated by the strong magnetic field group, while the transition region and corona are energized predominantly by the weak magnetic field group (Li $\&$ Feng 2022; Li 2024; Chen et al. 2025). Crucially, these heated atmospheric layers exhibit in-phase or anti-phase with the solar cycle, mirroring the temporal behavior of their governing magnetic field groups.

While the distinct layers of the solar atmosphere -the photosphere, chromosphere, transition region, and corona -are well-known, the reasons for this layered structure are less commonly explained. The phenomenon of anomalous heating (temperature distribution under atmospheric models) corresponds perfectly to this atmospheric stratification: the process of slow anomalous heating characterizes the chromosphere, while intense anomalous heating is localized to the corona.
In this sense, anomalous heating and atmospheric stratification are two manifestations of the same underlying problem, both arising from the influence of different groups of magnetic fields. However, the question of stratification is arguably more complex, as it must account not only for the temperature distribution but also for the sharp decrease in density in the upper atmosphere - a feature that may itself be shaped by magnetic field configurations.For instance, the umbrella-like (canopy) structure of the magnetic field may play a key role by impeding the outflow of charged particles, thereby causing the sharp drop in density observed in the upper atmosphere (Li $\&$ Feng 2022; Li 2024).

Solar wind is closely linked to atmospheric heating, both of which result from magnetic activity. Investigating the solar wind through the dual nature of magnetic fields may offer new avenues for understanding its dynamics. For example, low-latitude solar wind is typically slow-speed and aligns in phase with the solar cycle, while high-speed winds predominantly emerge at high latitudes and display an out-of-phase relationship with the solar cycle (Li 2016), somewhat reflecting this magnetic duality. Future research could focus on studying the long-term evolution of high- and low-speed solar wind across high- and low-latitude regions to further unravel their interconnected mechanisms.

Under the influence of solar magnetic fields, this dual nature extends even to the convection field itself (Chatterjee et al.  2017). Research has demonstrated that long-term variation in the mean scale of solar supergranules remains synchronized with the solar cycle - a correlation predominantly governed by magnetic activity within active regions. In contrast, quiet regions show suppressed anti-phase behavior, as their inverse relationship is overwhelmed by the in-phase signature detected across the Sun's surface (Chatterjee et al.  2017). This pattern directly corresponds to the temporal dynamics of their governing magnetic field groups, indicating a close relationship between magnetic fields at all scales and convection processes.

The solar cycle manifests in magnetic elements across the entire solar disk, though magnetic signals weaken significantly at high latitudes compared to low latitudes. Both large- and small-scale magnetic elements show correlations with the solar cycle, but key differences emerge: small-scale magnetic activity displays notably weaker signals and exhibits a phase offset relative to the large-scale cycle. This dual behavior suggests an interconnected relationship between magnetic elements of different scales, with the small-scale solar activity cycle acting almost like a ``boomerang" (or reflux) response to the large-scale cycle. Current theoretical models explain these phenomena through distinct mechanisms. Large-scale magnetic fields are governed by the global dynamo theory, which involves plasma flows and rotational shear in the Sun's interior, while small-scale fields are attributed to localized dynamo processes driven by near-surface convective turbulence (Vogler $\&$  Schussler 2007; Charbonneau 2020a; Karak 2023). Despite their apparent independence, the interaction between these two dynamo mechanisms- and whether they share underlying physical connections- remains a critical unresolved question in solar physics. The duality of magnetic fields implies that magnetism operates across multiple scales, and magnetic activity at high and low latitudes is intrinsically linked, highlighting the necessity of integrating low and high latitudes. This interconnectedness underscores the complexity of solar dynamics, where localized and global processes may influence one another in ways that are not yet fully understood. The duality framework may apply to solar-type stars, where surface shear and core-convection interactions shape magnetic cycles, and recent holistic magnetic braking models for stellar spin evolution support this cross-scale coupling (Bohm - Vitense 2007; Charbonneau 2020a).

Magnetic field measurements inherently contain observational uncertainties. However, this study reveals that all weak magnetic elements (WG) maintain a consistent phase relationship with the solar cycle, as do all strong magnetic elements (SG). Notably, both the intensity and number of WG elements show identical solar cycle dependence, with SG elements exhibiting the same internal consistency.
These findings demonstrate that the precise measured values of individual magnetic elements are not the determining factor. Rather, the critical distinction lies in their fundamental classification as either weak (WG) or strong (SG) magnetic elements.
When the number of data points used to calculate correlation coefficients exceeds 10,000, the minimum absolute correlation coefficient for both the strong and weak magnetic field groups-relative to solar cycles-remains above 0.31 (Table 2). This indicates that the solar cycle variations of these groups are  statistically reliable (p $<$ 0.001).
There may be cross-talk between the weak-field (WG) and strong-field (SG) groups. Correcting for this effect could enhance their observed relationship, since the long-term variations of WG and SG are anti-correlated. After removing this cross-talk contamination, the intrinsic WG signal should display even stronger solar-cycle variations. Consequently, the correlation coefficients in Table 2 would likely increase in magnitude. These long-term variations retain statistical significance across nearly all of the 180 measurement latitudes. Even when the magnetic field strength is divided into 150 intervals, the vast majority of these intervals still exhibit a statistically significant phase relationship consistent with the group's solar cycle behavior.

In general,  uncertainties in magnetic field measurements tend to attenuate the detected solar activity cycle signal, unless the uncertainties themselves contain the solar activity cycle signal. Currently, the prevailing scientific consensus attributes the approximately 11-year periodic signal to solar origins. The fact that the solar cycle signal remains detectable despite the presence of data uncertainty further strengthens confidence in its robustness and suggests that the identified signal is likely genuine. Our research reports detected solar activity cycle signals. The associated uncertainty merely complicates the detection process, it cannot negate the research.

The projection effect can interfere with detecting solar activity cycle signals. Crucially, however, the projection effect itself introduces variability on annual timescales and does not exhibit the 11-year periodicity of the solar cycle.  The detection of the solar cycle signal despite this interference lends it greater credibility. To further mitigate the impact of the projection effect, we employed two detection methods based on the number and magnetic intensity of magnetic elements.

Jin et al. (2015a) used data with different spatial resolutions and found that, while resolution had a minor impact  on correlation coefficients, it did not affect their overall conclusion. Our conclusion is consistent with the findings of Jin et al. (2015a), also indicating that spatial resolution has little significant impact on the detection of solar cycle signals.
Our findings align with those of Jin et al. (2011), who reported similar results at low latitudes using higher-precision daily magnetograms and different methodologies (though their analysis was restricted to within $60^{\circ}$ of the disk center). This consistency provides independent validation of our conclusions.

Finally, let's classify the four magnetic field components according to the traditional categories of solar magnetic fields. At low latitudes, Components I through IV correspond, in sequence, to network magnetic fields (NTFs), ephemeral region magnetic fields (ERFs), a combination of ERFs with a small contribution from active region magnetic fields (ARFs), and ARFs with a possible minor mixture of ERFs. That is, Group I consists of network magnetic fields (NTFs), while Group II is composed of active region magnetic fields (ARFs) and micro active region magnetic fields (ERFs).
At high latitudes, the observed magnetic fields are predominantly horizontal. Those with strengths greater than 18.37 G represent the horizontal magnetic fields of ephemeral regions. At both low and high latitudes, (micro) active regions  vary in phase with the solar cycle. This is consistent with the finding that filaments also follow the solar cycle at both low and high latitudes (Li et al. 2007).
At high latitudes, magnetic fields with strengths between 5.29 G and 18.37 G are identified as the observed horizontal network fields. These vary in anti-phase with the solar cycle, in agreement with Shiota et al. (2012). NTFs exhibit anti-phase behavior at all latitudes across the solar disk.
At high latitudes, magnetic fields weaker than 5.29 G are inferred to be likely horizontal internetwork fields.
These vary in phase with the solar cycle, consistent with Trelles Arjona et al. (2023). The reason why INFs are successfully observed by NSO/Kitt Peak at high latitudes - but not at low latitudes - is that INFs are highly inclined, with their horizontal flux density being approximately 8.7 times greater than their vertical magnetic flux density (Jin and Wang 2015b).
Due to the projection effect at high latitudes, magnetic elements appear more compact in visible area than they would at low latitudes (Shiota et al., 2012). Consequently, the same magnetic element exhibits a stronger line-of-sight magnetic flux when observed near the polar limb.
This strong horizontal flux in internetwork regions makes it difficult to distinguish between INFs and NTFs at high latitudes. Moreover, the intermixing with NTFs likely dilutes the solar cycle signal within internetwork fields, which may explain why many observational studies find no correlation between NTFs and the solar cycle."
Further, this can explain the weak signals reported by Shiota et al. (2012) in their Figure 5(e) and 5(f), as well as by Jin $\&$ Wang (2015b).
These interpretations are expected to be validated through future direct (face-up) observations of the solar polar regions.

{\bf Acknowledgements}\\
The authors are grateful to the anonymous referee for a careful reading and helpful comments.
The NSO/Kitt Peak data used here (synoptic magnetograms from Carrington rotations 1625 to 2007) are produced cooperatively by NSF/NOAO, NASA/GSFC, and NOAA/SEL, and they are used here and  can be publicly available from the NSO's web site, ftp:// nispdata.nso.edu/kpvt/ synoptic/.
Sunspot numbers (version 2) come from the Sunspot Index and Long-term Solar Observations (WDC-SILSO), Royal Observatory of Belgium, Brussels,
which can be downloaded from theSILSO's web site, https://www.sidc.be/SILSO/datafiles.
This work is supported by the National Natural Science Foundation of China (12373059 and 12373061), the Yunling-Scholar Project (the Yunnan Ten-Thousand Talents Plan),  the ``Yunnan Revitalization Talent Support Program" Innovation Team Project (202405AS350012), the project supported by the specialized research fund for state key laboratories, and the Chinese Academy of Sciences.

\label{lastpage}
\end{document}